\documentclass[preprint,amsmath,amssymb,aps]{revtex4-1}

\usepackage{graphicx}
\usepackage{color}
\usepackage{epsfig} 
\usepackage{pstricks}

\begin{document}

\title{Optoelectronic properties of zinc oxide: A first-principles investigation using the
Tran-Blaha modified Becke-Johnson potential}
\author{R. M. V. S. Almeida}
\affiliation{Instituto de F\'{\i}sica, Universidade Federal da Bahia, Campus Universitario de Ondina, 40210-340 Salvador, Bahia, Brazil}
\author{A. L. da Rosa}
\affiliation{Instituto de F\'{\i}sica, Universidade Federal de Goi\'{a}s, 74.690-900 Goi\^{a}nia, Goi\'{a}s, Brazil}
\author{J. S. de Almeida}
\email[]{jailton$_$almeida@hotmail.com}
\affiliation{Instituto de F\'{\i}sica, Universidade Federal da Bahia, Campus Universitario de Ondina, 40210-340 Salvador, Bahia, Brazil}
\date{\today}

\begin{abstract}
In this work we use density functional theory (DFT) to investigate the
influence of semi-local exchange and correlation effects on the
electronic and optical properties of zinc oxide. We find that the
inclusion of such effects using the Tran-Blaha modified Becke-Johnson
potential yields an excellent description of the electronic structure
of this material giving energy band gap which is systematically larger
than the one obtained with standard local functionals such as the
generalized gradient approximation. The discrepancy between the
experimental and theoretical band gaps is then significantly reduced
at a computational low cost. We also calculated the dielectric functions
of ZnO and find a violet shift to the absorption edge which is in 
good agreement with experimental results.
\end{abstract}

\maketitle   
\section{Introduction}
\label{sec:s1} 

Zinc oxide (ZnO) is a wide band gap semiconductor with promising
application in laser diodes, light emitting diodes, transparent
electrodes and gas sensors \cite{Morkoc:09}. Although there has been a
great progress in samples fabrication, controlling the conductivity in
ZnO is still a challenge due to the presence of impurities and
defects\cite{Janotti:07,Lany:10}.  From the theoretical side, the
correct description of the band gap is of paramount importance for the
understanding of impurity and defect states in semiconductors. It is a
common understanding that the use of local exchange-correlation
functionals wrongly describe the ZnO band gap. This can also lead to
misleading conclusions on the location of the impurity and defect
states \cite{Lany:08,Lany:09,Lany:10,Janotti:09,Sarsari:13}. The
experimental band gap of ZnO is 3.4 eV \cite{madelung:parameters}
while the calculated value using GGA-PBE \cite{Perdew:96} yield
0.7-0.9 eV\,\cite{Lany:08,Lany:09,Janotti:09}. There are different
schemes available to improve the band gap, one of the most common
being hybrid functionals\,\cite{HSE:06,PBE0}. More sophisticated
approaches, such as the GW method\,\cite{Baym:61,Baym:62,Hedin:65} has
become the state of the art but it is still computationally cost. It
has been shown that partially or fully self-consistent schemes
\cite{Shishkin:07,Rinke:05,Thygesen:2013,Louie:2014,Lany:10,Gori:10},
in which either Green's function G or both the Green's function and
the dielectric matrix are updated can improve the agreement with
experiments. Although for small cells the calculations are feasible
when treating larger systems required to investigate defects these
schemes are computationally much more demanding compared to LDA or GGA
and for many systems the GW approximation is still
prohibitive. Therefore one looks always for alternative methods which
are computationally at similar costs as local/semi-local DFT
calculations and at the same time could improve band gaps and
description of energy level positions.

Recently proposed Tran-Blaha modified version of the Becke-Johnson
potential (TB-mBJ) \cite{Blaha:09} has proved to be a successful method for
accurate band gaps of semiconductors and insulators \cite{Almeida:13,Rai:14,Tran:07,Haq:13,Dixit:12}. Furthermore, it
has a computational cost as the LDA or GGA. For ZnO, it has been shown to lead 
to band gaps close to the experimental values \cite{Haq:13,Dixit:12,madelung:parameters}. 

In this work, we use the Tran-Blaha modified Becke-Johnson (TB-mBJ)
potential \cite{Blaha:09} to investigate the electronic structure and
optical response of ZnO. We found that TB-mBJ potential greatly
improves the description of the electronic structure of ZnO compared
to GGA, but still has several drawbacks with respect to GW
calculations. 

\section{Computational details} 

\begin{figure*}
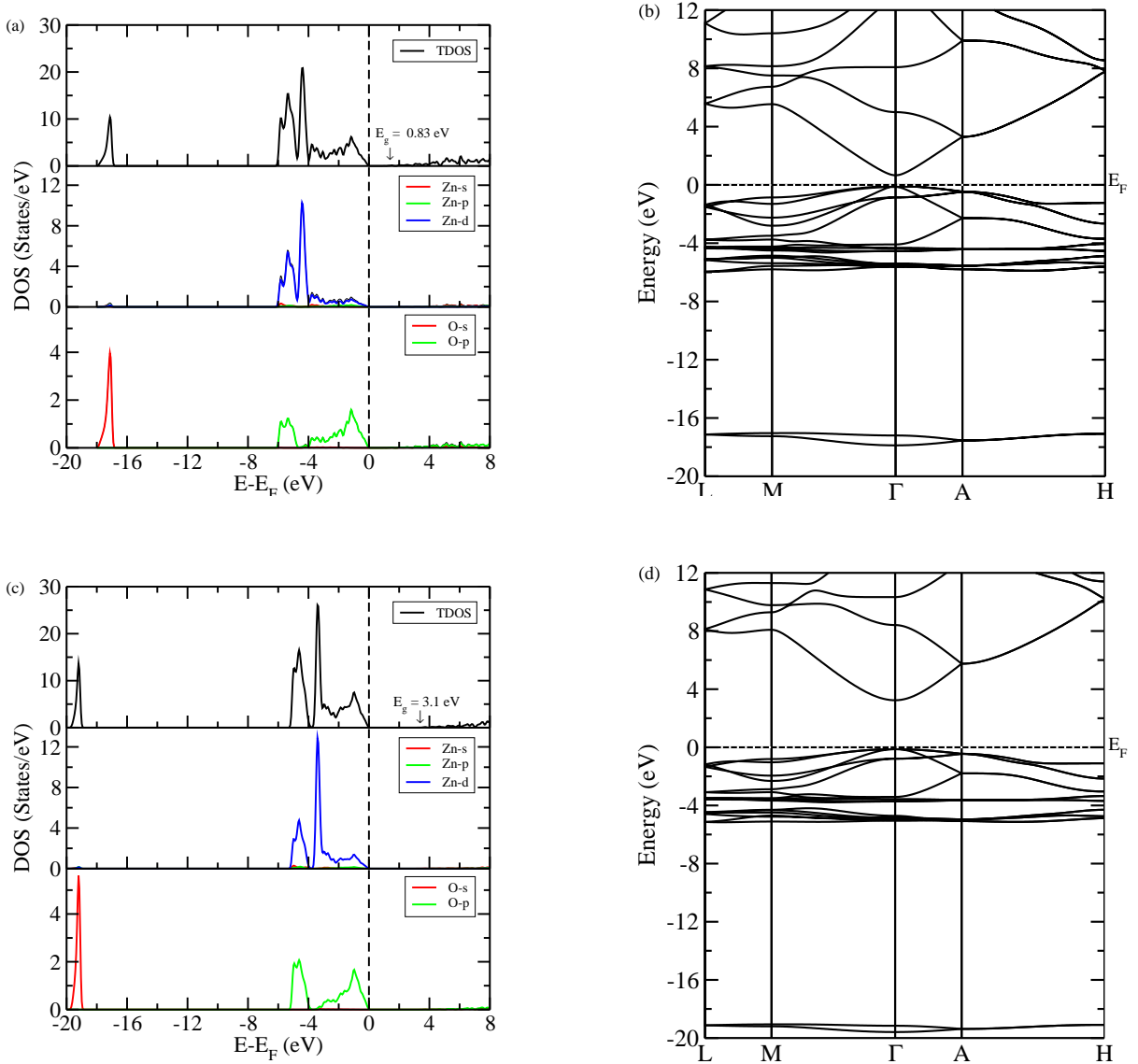

\pspicture(0,0)(16,15.2)
\rput[bl](0,8){\epsfig{file=fig1a.eps,width=7cm}}
\rput[bl](9,8){\epsfig{file=fig1b.eps,width=7cm}}
\rput[bl](0,0){\epsfig{file=fig1c.eps,width=7cm}}
\rput[bl](9,0){\epsfig{file=fig1d.eps,width=7cm}}
\endpspicture
\caption{Total and partial density of states (DOS) (a) and band structure (b) of ZnO 
with GGA-PBE functional (upper panel). Total and partial DOS (c) and band structure (d) of ZnO using TB-mBJ potential (lower panel).}
\label{fig:um}
\end{figure*}

To investigate the optoelectronic properties of ZnO, we have used
density-functional theory within the projected augmented wave (PAW)
method \cite{Bloechel:94} and the GW technique as implemented in the
Vienna Ab initio Simulation Package (VASP) \cite{Kresse:99}. The
exchange and correlation potential was described using the generalized
gradient approximation (GGA) in the Perdew, Burke, and Ernzerhof (PBE)
parametrization \cite{Perdew:96}, the Tran-Blaha modified
Becke-Johnson (TB-mBJ) potential \cite{Blaha:09} and the GW
method. The TB-mBJ potential is a modified version of the
Becke-Johnson potential \cite{Becke:06} used to improve band gaps
obtained by the conventional DFT-based methods. The TB-mBJ potential
can be written as

\begin{eqnarray}\label{eq01}
\upsilon_{x,\sigma}^{TB-mBJ}(\mathbf{r}) = c\upsilon_{x,\sigma}^{BR}(\mathbf{r})+(3c-2)\frac{1}{\pi}\sqrt{\frac{5}{6}}\sqrt\frac{\tau_{\sigma}(\mathbf{r})}{\rho_{\sigma}(\mathbf{r})}
\end{eqnarray}
where $\rho_{\sigma}= \sum^{N_{\sigma}}_{i=1}\left|\psi_{i,\sigma}
\right|^{2}$ is the electron density,
$\tau_{\sigma}=\frac{1}{2}\sum^{N_{\sigma}}_{i=1} \nabla
\psi^{\ast}_{i,\sigma} . \nabla \psi_{i,\sigma} $
is the kinetic-energy
and $\upsilon_{x,\sigma}^{BR}$ is the Becke and Roussel potential
 \cite{Becke:06}. The $c$ parameter value showed in Eq. 1 is computed with
as

\begin{eqnarray}\label{eq02}
c = \alpha + \beta \left( \frac{1}{V_{cell}}\int_{cell}\frac{\left|\nabla{\rho _\sigma (r')}\right|}{\rho_\sigma(r')}d^{3}r'\right)^{\frac{1}{2}}
\end{eqnarray}

where $V_{cell}$ means the unit cell volume, $\alpha =-0.012$ and
$\beta=1.023$ bohr$^{\frac{1}{2}}$ are parameters fitted according to
experimental values for several materials\,\cite{Blaha:09}. It is
important to mention that TB-mBJ is a potential-only functional which
means there is no corresponding TB-mBJ exchange-correlation
energy. This fact, therefore, leads to the impossibility of using
TB-mBJ to compute Hellmann-Feynman forces and to compare total
energies. The PAW potentials with the valence states 2{\it s} and
2{\it p} for O atom and 3{\it d} and 4{\it s} for Zn atom and a basis
set up to a kinetic energy cutoff of 450 eV have been
used. Integration over the Brillouin zone was performed using a
$9\times 9\times 7$ Monkhorst-Pack {\bf k}-points grid. In the optical
properties calculations, however, the irreducible Brillouin zone (IBZ)
has been sampled with about 500 \textbf{k}-points. All the
calculations have been carried out until the Hellmann-Feynman forces
become smaller than $10^{-3}$ eV/{\AA} and the total energies
converged to below $10^{-4}$ eV with respect to the Brillouin zone
integration.

As a matter of comparison we have also performed calculations for the
dielectric function in the GW approximation.  Several approximations
lead to different results, such as number of bands the
exchange-correlation potential for the starting wave function as well
as the use of approximate models for the screening, like plasmon pole
approximations.  Depending on the starting functional and the details
of the calculation, values between 2.1 and 3.6 eV are obtained for
G$_0$W$_0$
\cite{Shishkin:07,Sarsari:13,Thygesen:2013,Louie:2014,Friedrich:11,Shishkin:07,Shih:10},
between 2.54–3.6 for GW$_0$
\cite{Sarsari:13,Thygesen:2013,Louie:2014,Friedrich:11,Shishkin:07,Shih:10}
and between 3.2–4.3 for GW \cite{Sarsari:13,Usuda}. For the GW$_0$
calculations, a cutoff of 100 eV for the response functions, as well
as 1024 bands have been employed. We obtained 3.2\,eV in fair
agreement with optical experiments\cite{madelung:parameters} and other
calculations within the same
approach.\,\cite{Sarsari:13,Thygesen:2013,Louie:2014,Friedrich:11,Shishkin:07,Shih:10}. The
position of the Zn-3d states has also been corrected to -7.0\,eV,
closer to the experimental values \cite{Shishkin:07} and in agreement with
previous GW$_0$ calculations\,\cite{Shishkin:07,Thygesen:2013}.

\section{Electronic Properties}

ZnO has a wurtzite hexagonal crystal structure that belongs to the
$P6_3mc$ space group with the unit cell containing two Zn cations and
two O anions. The lattice parameters are the second nearest neighbor
distances and each Zn atom is coordinated by four O atoms which are
located at the corners of a slightly distorted tetrahedron.  The Zn-O
distance along the {\it c} axis is about 1.90 \AA \ which is a little
shorter than 1.98 \AA \ of its counterpart perpendicular to the {\it
  c} axis.

To investigate the optoelectronic properties of ZnO, we firstly
calculated the optimized crystal structure of ZnO using the GGA-PBE
functional. Thereafter we computed the electronic structure and
dielectric functions of ZnO with both GGA-PBE functional and TB-mBJ
potential. We found that the GGA-PBE optimized lattice constants for
ZnO crystal are $a = 3.291$ \AA \ and $c = 5.266$ \AA \, which are in
agreement with the experimental values of $a = 3.250$ \AA\ and $c =
5.207$ \AA \cite{Kisi:89}.

The electronic structure of the ZnO has been calculated with GGA-PBE
functional and TB-mBJ potential and they are shown in
Fig.\ref{fig:um}, respectively. The Fermi-level (dotted line) has been
specified to be zero in this paper.  The upper panel of
Fig.\ref{fig:um} presents the results for DOS and band structure of
ZnO using GGA-PBE functional while the lower panel those calculated
with TB-mBJ potential. The DOS spectrum (Fig.\ref{fig:um} (a))
possesses three major features in the valence band. The lowest part of
it from -18 to -16.8 eV is mainly contributed by O 2{\it s}
states. The second subband in the energy range between -6 and -4 eV
comes mostly from contributions of Zn 3{\it d} states whereas the
upmost valence subband located between -4 and 0 eV is mainly from O
2{\it p} states.  We note that there is also orbital hybridization
between Zn 3{\it d} and O 2{\it p} states from -6 to 0 eV. The
conduction band originates mainly by the Zn 4{\it s} states and the O
2{\it p} states. Moreover, the valence and conduction bands are
separated by the energy band gap which in our GGA-PBE calculations
yields a value of 0.83 eV and this result is consistent with previous
calculations \cite{Lany:08,Janotti:09}.

It is known that DFT calculations using standard
functionals give a very small band gap as compared with experimental
values. This effect in ZnO is further enhanced due to the
underestimation of the repulsion between the Zn 3{\it d} states and
the conduction band \cite{Janotti:09}, which induces to a significant
hybridization of the O 2{\it p} and Zn 3{\it d} levels. In this case,
the resulting overly large level of repulsion between the Zn 3{\it d}
states and valence bands pushes the valence band maximum up.

In order to overcome the band gap problem and have a good description
of the electronic structure of ZnO, we applied the TB-mBJ potential on
top of the GGA-PBE calculations. We found that the calculated DOS with
TB-mBJ method possesses approximately similar features to those of
pure ZnO with functional GGA-PBE. However in the TB-mBJ DOS
(Fig.\ref{fig:um}(c)) there is a shift of O 2{\it s} states towards
lower energy and we note that TB-mBJ potential gives an improved band
gap value of 3.10 eV which is remarkable close to the experimental
value of about 3.40 eV \cite{madelung:parameters}.

\begin{figure}[!ht]
\pspicture(0,0)(8,7)
\rput[bl](0,0){\epsfig{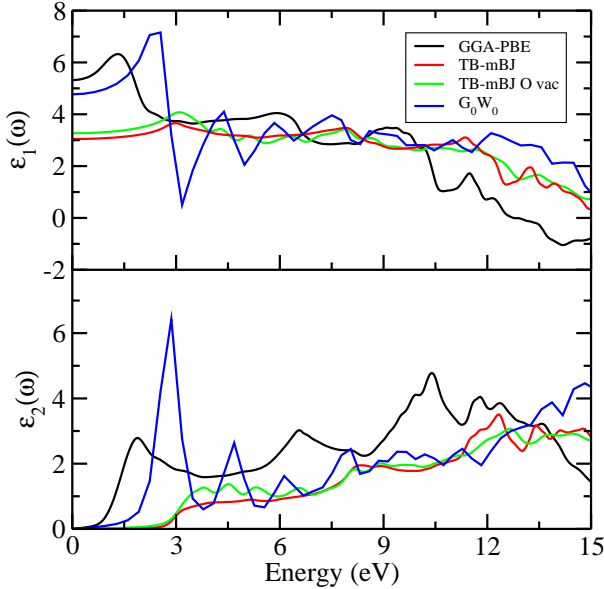}}
\endpspicture
\caption{Average of real (upper panel) and imaginary (lower panel) parts of the dielectric function of ZnO bulk 
crystal calculated using the GGA-PBE (black line) and TB-mBJ (red line) exchange-correlation functionals and using
G$_0$W$_0$ method (blue line). Results of the dielectric functions for ZnO with a neutral oxygen vacancy obtained using
TB-mBJ (green line) method are also presented.}
\label{fig:dois}
\end{figure}

The band structure along high symmetry lines in the hexagonal
Brillouin zone (BZ) using GGA-PBE functional is shown in
Fig.\ref{fig:um}(b). It is evident from the figure that ZnO exhibits a
direct band gap with both valence band maximum and conduction band
minimum located at BZ center at $\Gamma$ point. The lowest two bands
(occurring around -16.6 eV) correspond to O 2{\it s} levels. The next
set of ten bands (occurring around -6 eV) are due to Zn 3{\it d}
levels. Then six bands from -6 to 0 eV correspond to O 2{\it p}
bonding states. The first two conduction bands are predominantly
formed by Zn 4{\it s} states. In Fig.\ref{fig:um}(d), we present the
band structure of ZnO calculated with TB-mBJ potential. We observe a
band narrowing in the upper valence band of about 1 eV leading to more
localized electrons in the energy range of -6 to 0 eV when compared to
GGA-PBE band structure (Fig.\ref{fig:um}(b)). Additionally, as
mentioned before, we obtained a band gap of 3.10\,eV with TB-mBJ
potential which is much improved relative to the GGA-PBE functional
and is in good agreement with the experimental value
\cite{madelung:parameters}.

\section{Optical Properties}

The optical properties of ZnO have been investigated through the
imaginary and real parts of the dielectric function. The imaginary
part of the dielectric function is calculated directly from the
electronic structure through the joint density of states and the
matrix elements of the momentum, \textbf{p}, between occupied and
unoccupied eigenstates according to:

\begin{eqnarray}
\epsilon_{2}^{ij}(\omega)&=& {4\pi^2 e^2 \over \Omega m^2 \omega^2}
\sum_{{\bf k} n n^{\prime}}
\bigl\langle {\bf k} n \big | p_{i} \big | {\bf k}
n^{\prime} \bigr\rangle
\bigl\langle {\bf k} n^{\prime} \big | p_{j} \big | {\bf k}
n \bigr\rangle \times\nonumber\\
&& \times f_{{\bf k}n}\, \bigl(1 - f_{{\bf k} n^{\prime}}\bigr) \,
\delta\bigl( E_{{\bf k} n^{\prime}} - E_{{\bf k} n} - \hbar \omega
\bigr).\ \
\end{eqnarray}

In this equation, {\it e} is the electron charge, {\it m} the electron mass,
$\Omega$ is the volume of the crystal, $f_{\bf k n}$ is the Fermi
distribution function and $\big |\bf k {\it n} \bigr\rangle$ is the 
crystal wave function corresponding to the ${\it n^{th}}$ eigenvalue
${\it E}_{{\bf k}n}$ with crystal wave vector {\bf k}. The real part
of the dielectric function is calculated from $\epsilon_{2}$ via the
Kramers-Kronig transformation

\begin{equation}
\epsilon_{1}(\omega) 
 =1+{2\over{\pi}}\int_{0}^{\infty}{{d\omega^{\prime}
\epsilon_{2}(\omega^{\prime})}\Big({\omega^{\prime}\over \omega^{\prime2}-\omega^2}\Big)}\\, 
\end{equation}
where the integral is evaluated setting the frequency cut-off to be several times larger than the frequency range
in order to produce accurate results.


In anisotropic materials, dielectric properties must be described by
the dielectric tensor. In the case of ZnO, the tensor components can
be reduced to only two independent components $\epsilon_{\rm xx}$ =
$\epsilon_{\rm yy}$ $\neq$ $\epsilon_{\rm zz}$. For materials with weak
anisotropy, it has been shown \cite{(Calliari} that the
dielectric function can be replaced by an average
$\epsilon = \frac{2\epsilon_{\rm xx} + \epsilon_{\rm zz}}{3}$.

In Fig. \ref{fig:dois} we show the average of the parallel and
perpendicular directions of light
polarization for $\varepsilon_{1}(\omega)$ and $\varepsilon_{2}(\omega)$ 
of ZnO using GGA-PBE (black line), TB-mBJ (red line) and G$_0$W$_0$ (blue line)
methods. We also show the average of the dielectric function of
ZnO with a neutral oxygen vacancy calculated using the
TB-mBJ (green line) method. It is worth to mention that $\varepsilon_{2}(\omega)$ 
correspond to the
experimental light absorption such as observed in electron-energy loss
spectroscopy (EELS) \cite{Rodriguez:2011}. GGA-PBE leads to a too
small absorption edge when compared to experimental results
\cite{Jellison:98,Gori:10}. We performed calculations applying the
TB-mBJ potential as shown in the red curve. The use of TB-mBJ
correction gives a violet shift to the optical band gap of ZnO and it
nicely reproduces the experimental \cite{Jellison:98,Gori:10}
absorption edge of ZnO at low computational cost.

The $\varepsilon_{2}(\omega)$ spectrum of ZnO shows a strong optical
anisotropy and it can be roughly divided into three main regions,
namely, the low-energy which is the region below 5 eV, the
middle-energy one from 5 to 10 eV and the high-energy region above 10
eV. For GGA-PBE calculations, the first peak of $\epsilon_2(\omega)$
is located at about 2 eV but it appears as a shoulder at about 3 eV
for TB-mBJ results. These features are formed mainly due to electronic
transitions at the $\Gamma$ point occurring between O 2{\it p} and Zn
4{\it s} orbitals. In the middle energy region, there is one peak
located at 6.6 eV for the GGA-PBE and it appears less prominent at 8.2
eV for the TB-mBJ. These peaks in the middle-energy region are mainly
derived from the transition at M and L points between the Zn 3{\it d}
and O 2{\it p} orbitals. The high-energy region exhibits two principal
peaks, around 10 eV and 12 eV for the GGA-PBE, and 12.3 eV and 14.5 eV
for the TB-mBJ. The peaks correspond mainly to the transitions from Zn
3{\it d} and O 2{\it s} orbitals.  The TB-mBJ calculations for
$\varepsilon_{2}$ show a good overall agreement with the experimental
results, but still fails compared to G$_0$W$_0$ calculations. The strong
absorption peak around xx eV is very broad with TB-mBJ, which is
probably due to the wrong description of the charge density. 

The GW$_0$ shown in blue gives a different picture, with a strong
absorption at 3.15 eV.  The results for the real part, $\epsilon_1(\omega)$, of the dielectric
function are consistent with the GGA-PBE and TB-mBJ calculations of
$\epsilon_2(\omega)$. By looking at the upper panel of
Fig.\ref{fig:dois}, it is possible to verify that the optical response
of ZnO in the transparent region for GGA-PBE is narrowed compared to
the TB-mBJ results due to the shortness of the band gap as previously
discussed.

\section{Conclusion}

In this paper we have investigated the electronic and optical
properties of zinc oxide by using density function theory including
semilocal exchange-correlation potentials and the GW$_0$ method. We
found that the band gap and absorption edge of ZnO are in good
agreement with experimental data. However, the intensity of the peaks
does not agree well with GW$_0$ due to the lack of a better
description of screening effects. The inclusion of screening effects
are primordial to achieve a better agreement with experimental
results. Still, the present results strongly support the conclusion
that TB-mBJ potential greatly improves the band gaps and electronic
structure of simple semiconductors and insulators.

\section{acknowledgement}
We are thankful for the financial support from the Brazilian agencies
CNPq, CAPES, and FAPESB. A.L.R would like to thank German Science Foundation (DFG) under the program FOR1616.


\providecommand{\WileyBibTextsc}{}
\let\textsc\WileyBibTextsc
\providecommand{\othercit}{}
\providecommand{\jr}[1]{#1}
\providecommand{\etal}{~et~al.}

\end{document}